\documentclass[journal]{IEEEtran}

\usepackage{amsmath,amsfonts}
\usepackage[ruled]{algorithm2e}
\usepackage{array}
\usepackage{textcomp}
\usepackage{stfloats}
\usepackage{url}
\usepackage{verbatim}
\usepackage{graphicx}
\usepackage{cite}
\usepackage{amssymb}
\usepackage{xcolor}
\usepackage{comment}
\usepackage{multicol}
\usepackage{multirow}
\usepackage{subfigure}
\usepackage{setspace}
\usepackage{psfrag}
\usepackage{stmaryrd}
\usepackage{mathrsfs}
\usepackage{epstopdf}
\usepackage{enumerate}
\usepackage{booktabs}  
\usepackage{color}
\usepackage{dsfont,bm}
\usepackage{epsfig}
\usepackage{balance}
\usepackage{enumitem}
\usepackage{amsthm}

\newtheorem{theorem}{Theorem}

\newtheorem{lemma}{Lemma}

\newtheorem{corollary}{Corollary}

\allowdisplaybreaks[4]
\begin{document}

\title{Dynamic Hybrid Beamforming Design for Dual-Function Radar-Communication Systems}

\author{\IEEEauthorblockN{
    Bowen Wang,~\IEEEmembership{Graduate Student Member,~IEEE,}
	Hongyu Li,~\IEEEmembership{Graduate Student Member,~IEEE,} \\and
	Ziyang Cheng,~\IEEEmembership{Member,~IEEE} 
    \vspace{-2.5em}
}
\thanks{Copyright (c) 2015 IEEE. Personal use of this material is permitted. However, permission to use this material for any other purposes must be obtained from the IEEE by sending a request to pubs-permissions@ieee.org.}
\thanks{Manuscript received 11 September 2022; revised 26 March 2023, and 7 July 2023; accepted 2 September 2023; Date of publication 11 September 2023.
The work of B. Wang and Z. Cheng was supported in part by the National Natural Science Foundation of China under Grants 62371096, 62231006 and 62001084, and in part by Sichuan Science and Technology Program   2023NSFSC1385.
The review of this article was coordinated by Prof. Byonghyo Shim. \textit{(Corresponding author: Ziyang Cheng)}.}
\thanks{B. Wang and Z. Cheng are with the School of Information and Communication Engineering, University of Electronic Science and Technology of China, Chengdu 611731, China. (e-mail: bwwang@std.uestc.edu.cn, zycheng@uestc.edu.cn).}
\thanks{H. Li is with the Department of Electrical \& Electronic Engineering, Imperial College London, London SW7 2AZ, U.K. (e-mail: c.li21@imperial.ac.uk).}
}



\maketitle

\begin{abstract}
This paper investigates dynamic hybrid beamforming (HBF) for a dual-function radar-communication (DFRC) system, where the  DFRC base station (BS) simultaneously serves multiple single-antenna users and senses a target in the presence of multiple clutters.
Particularly, we apply a HBF architecture with dynamic subarrays and double phase shifters in the DFRC BS.
Aiming at maximizing the radar mutual information, we consider jointly designing the dynamic HBF of the DFRC system, subject to the constraints of communication quality of service (QoS), transmit power, and analog beamformer.
To solve the complicated non-convex optimization, an efficient alternating optimization algorithm based on the majorization-minimization methods is developed.
Simulation results verify the advancement of the considered HBF architecture and the effectiveness of the proposed design method.
\end{abstract}
\vspace{-0.3em}

\vspace{-1em}
\section{Introduction}

Dual-function radar-communication (DFRC), as an integration of radar and communication functionality, has been envisioned as a promising technology for intelligent vehicular networks (IVNs) \cite{Liu2022Integrated}.
Nowadays, by integrating millimeter wave (mmWave) DFRC and massive multiple-input multiple-output (MIMO), the DFRC can achieve high-precision sensing while guaranteeing high-throughput communications, which has emerged as a potential enabler for IVNs \cite{Liu2022Integrated,zhang2021overview,zhou2022integrated}.
However, implementing mmWave massive MIMO DFRC systems with fully-digital beamforming architecture  is impractical due to the substantial power consumption and prohibitive hardware cost of mmWave RF components.
As a cost-effective alternative, the hybrid beamforming (HBF) architecture, dividing signal processing into high-dimensional analog beamforming (ABF) and low-dimensional digital beamforming (DBF), is envisioned as a good trade-off between system performance and hardware complexity \cite{ahmed2018survey}.

Initial works of HBF design for DFRC systems are limited to fixed connected ABF architectures \cite{qi2022hybrid,cheng2021hybrid,liu2020joint,cheng2021double}.
Specifically, the authors in \cite{qi2022hybrid} study the HBF design for the DFRC system with the fully-connected (FC) architecture, where the beampattern matching error is minimized subject to communication quality of service (QoS).
While the FC HBF architecture enables satisfactory performance, using a large number of phase shifters (PSs) may still cause high hardware complexity and power consumption. 
To facilitate the practical realization of HBF architecture, a fixed sub-connected HBF based DFRC system, where a single PS (SPS) is used to realize each RF chain-antenna connection, is designed in \cite{cheng2021hybrid}.
This architecture to a large extent reduces the number of PSs, but at the expense of non-negligible performance loss.
To make a balanced trade-off between performance and the number of PSs, the authors in \cite{cheng2021double} consider designing DFRC with double PS (DPS) based fixed sub-connected HBF architecture, where each RF chain-antenna connection is realized by DPS. 
Results demonstrate that HBF with DPS can significantly improve the performance over that with SPS by doubling the number of PSs.

The aforementioned works on DFRC systems \cite{liu2020joint,qi2022hybrid,cheng2021hybrid,cheng2021double} are restricted to fixed sub-connected HBF architectures, which limits the flexibility of ABF and further causes notable performance degradation. 
Recent communication literature unveils that the dynamically sub-connected architecture \cite{li2020dynamic,yu2018hardware,yu2019doubling}, which adaptively partitions all transmit antennas into several subarrays and maps each RF chain to one of them, has benefits on both spectral efficiency and energy efficiency in wireless communications compared with fixed sub-connected HBF architectures.
This motivates us to adopt this dynamically sub-connected HBF architecture at the DFRC transmitter to enhance the communication and sensing performance while maintaining affordable power consumption.

The contributions of this work are summarized as follows.

\textit{First}, we propose a novel DFRC system equipped with an energy-efficient dynamic HBF architecture.
Specifically, the ABF is realized by a switch network (SWNet) and a DPS based PS network (PSNet), which results in a dynamic connection between RF chains and transmit antennas. 

\textit{Second}, we formulate a dynamic HBF design problem to maximize the radar mutual information (RMI), subject to the constraints of QoS, transmit power, and ABF.
To solve this problem, we present an efficient alternating optimization (AO) algorithm based on the majorization-minimization method. 

\textit{Third}, we present simulation results to verify the effectiveness of the proposed algorithms and unveil the superior of the considered dynamically sub-connected HBF architecture over conventional fixed sub-connected architectures in terms of communication and radar sensing performance.

\textit{Notation:} 
$ (\cdot)^T $ and $ (\cdot)^H $ represent the  transpose and conjugate transpose operators. 
$ {\rm Vec}[\bf A] $   and ${\rm Tr}[{\bf A}]$ denote the vectorization  and trace of $ \bf A $, respectively.   
The operator ${\bf{A}}[ {i,j}]$ represents the $( i,j) $-th element of the matrix $\bf A$.
${\| {\; \cdot \;} \|_F}$ and ${\| {\; \cdot \;} \|_0}$ denote the Frobenius norm and 0-norm, respectively.

\section{System Model and Performance Metric}

\begin{figure}
    \centering
    \includegraphics[width=1\linewidth]{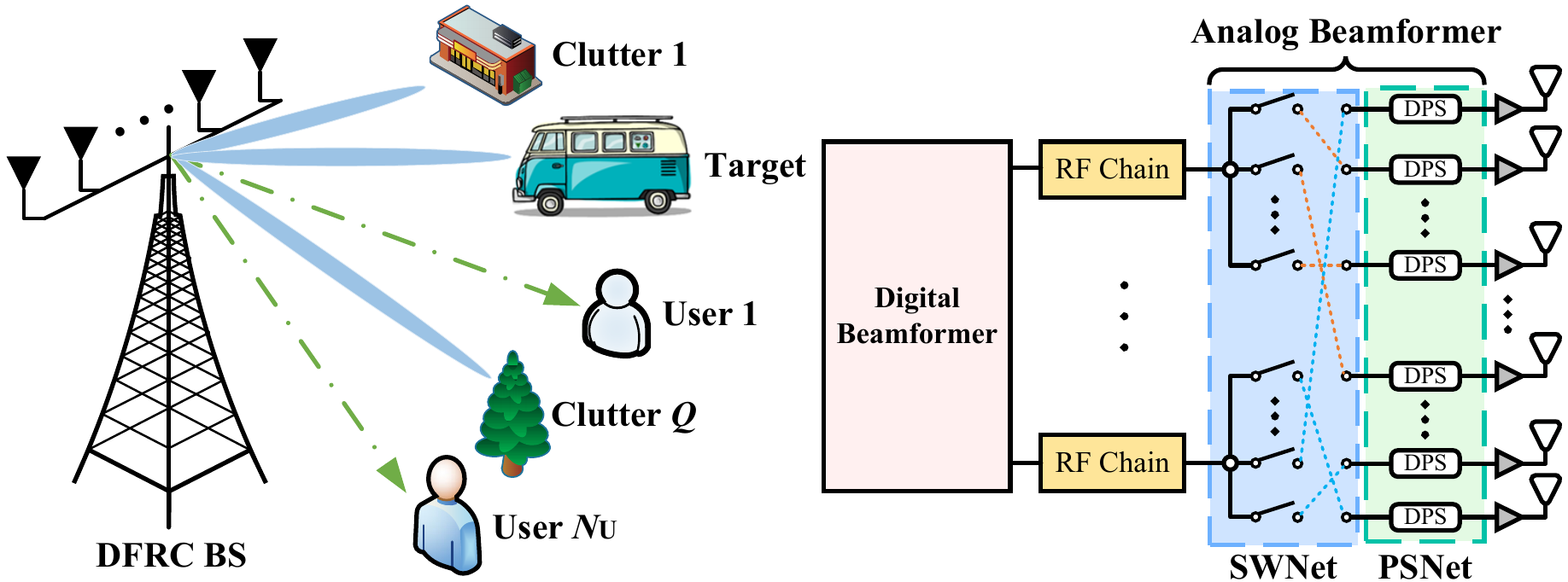}
    \vspace{-1em}
    \caption{Left: Diagram of the considered DFRC system. Right: Diagram of the dynamically sub-connected HBF architecture.}
    \vspace{-1.5em}
    \label{fig:sys0}
\end{figure}

In this paper, we consider a typical vehicle-to-infrastructure (V2I) scenario as shown in the left side of Fig. \ref{fig:sys0}, where a DFRC base station (BS) with $N_{\rm T}$ transmit antennas probs waveform toward a vehicle target in the presence of strong signal-dependent clutters and simultaneously provides downlink communication service to $N_{\rm U}$ single-antenna users.
The DFRC BS has $N_{\rm R}$ receive antennas to receive the scatter back waveform from the target and clutters and performs the target sensing task.
Both the transmit and receive antennas are arranged as the spaced uniform linear array (ULA).

\vspace{-1em}
\subsection{System Model}
Utilizing the HBF technique, the data stream ${\bf{s}}\left[ l \right] = {\left[ {{s_1}\left[ l \right], \cdots ,{s_{{N_{\text{U}}}}}\left[ l \right]} \right]^T} \in {\mathbb{C}}^{N_{\rm U}}$ at $l^{\rm th}$ time slot in the baseband are firstly processed by a low-dimensional DBF ${\bf F}_{\rm D} \in {\mathbb{C}}^{N_{\rm RF}\times {N_{\rm S}}}$.
Without loss of generality, we assume that the data symbols intended to different users are mutually independent and data streams between different time slot are uncorrelated, i.e, ${\mathbb E}\left\{ {{\bf{s}}\left[ l \right]{{\bf{s}}^H}\left[ l \right]} \right\} = {{\bf{I}}_{{N_{\text{U}}}}},\forall l$ and ${\mathbb E}\left\{ {{\bf{s}}\left[ l \right]{{\bf{s}}^H}\left[ \ell  \right]} \right\} = {{\bf{0}}_{{N_{\text{U}}}}},\forall l \ne \ell $.
Following the digitally precoded sequences, the DFRC BS then employ the ABF ${\bf F}_{\rm A} \in {\mathbb{C}}^{N_{\rm T}\times {N_{\rm RF}}}$ to map the RF signals form ${N_{\rm RF}}$ RF chains to $N_{\rm T}$ antennas.
Then, the transmit signal can be expressed as
\vspace{-0.5em}
\begin{equation}
	{\bf{x}}\left[ l \right] = {{\bf{F}}_{\text{A}}}{{\bf{F}}_{\text{D}}}{\bf{s}}\left[ l \right] = {{\bf{F}}_{\text{A}}}\sum\limits_{u \in {\mathcal U}} {{{\bf{f}}_{{\text{D}},u}}{s_u}\left[ l \right]} ,
	\label{eq:1}
    \vspace{-0.5em}
\end{equation}
where ${{\bf{f}}_{{\text{D}},u}}$ is the $u^{\rm th}$ column of digital beamformer ${\bf F}_{\rm D}$, i.e., ${{\bf{F}}_{\text{D}}} = \left[ {{{\bf{f}}_{{\text{D}},1}}, \cdots ,{{\bf{f}}_{{\text{D}},{N_{\text{U}}}}}} \right]$ and ${\mathcal{U}} = \{ {1, \cdots ,{N_{\text{U}}}} \}$.

In this paper, we employ a dynamically sub-connected ABF architecture in which ${\bf F}_{\rm A}$ is implemented using the SWNet and PSNet as shown in the right side of Fig. \ref{fig:sys0}.
Specifically, each element of PSNet is implemented by the DPS, where two PSs combine in parallel as illustrated in \cite{yu2018hardware,yu2019doubling,bogale2016number}.
The entry of each DPS is mathematically modeled as ${{\bf{F}}_{\text{A}}}\left[ {i,j} \right] = {e^{\jmath \phi _{i,j}^1}} + {e^{\jmath \phi _{i,j}^2}} = {A_{i,j}}{e^{\jmath {\phi _{i,j}}}}$, indicating that the beamforming coefficient is controlled by ${\phi _{i,j}^1}$ and ${\phi _{i,j}^2}$.
Moreover, the feedback link between the RF chain and antenna is controlled by SWNet to achieve dynamic mapping and diversity.
As for the mapping scheme, we assume there is no inter-subarray overlap, that is, each antenna is connected to only one RF chain, yielding the constraint ${\left\| {{{\bf{F}}_{\text{A}}}\left[ {i,:} \right]} \right\|_0} = 1,\forall i$.
Thus, the ABF should satisfy
\begin{equation}
	{{\bf{F}}_{\text{A}}} \in {\mathcal S} = \left\{ {{{\bf{F}}_{\text{A}}}:\left| {{{\bf{F}}_{\text{A}}}\left[ {i,j} \right]} \right| \le 2,{{\left\| {{{\bf{F}}_{\text{A}}}\left[ {i,:} \right]} \right\|}_0} = 1,\forall i,j.} \right\} .
	\notag
\end{equation}

Transmit signal \eqref{eq:1} is the dual-function signal utilized to perform radar sensing and communication tasks. 
Therefore, our objective is to efficiently design the transmit HBF so as to obtain desirable DFRC performance. 
In the following subsections, we will elaborate on the communication and radar metrics, and formulate the optimization problem.

\vspace{-1em}
\subsection{Metric for Communication Performance}
In this subsection, we introduce the metric to evaluate the communication performance.
Specifically, the DFRC BS transmits $N_{\rm S}$ independent data streams to $N_{\rm U}$ downlink users in each time slot.
After propagating through the environment, the received signal at $u^{\rm th}$ user in $l^{\rm th}$ time slot is given by
\begin{equation}
	\begin{aligned}
		 &{y_u}\left[ l \right] = {\bf{h}}_u^H{\bf{x}}\left[ l \right] + {n_{c,u}}\left[ l \right] \\
		& = {\bf{h}}_u^H{{\bf{F}}_{\text{A}}}{{\bf{f}}_{{\text{D}},u}}{s_u}\left[ l \right] + \sum\limits_{v \in {\mathcal U}, v \ne u} {{\bf{h}}_u^H{{\bf{F}}_{\text{A}}}{{\bf{f}}_{{\text{D}},v}}{s_v}\left[ l \right]}  + {n_{c,u}}\left[ l \right],
	\end{aligned}
	\label{eq:2}
\end{equation}
where ${n_{c,u}}\left[ l \right] \sim {\mathcal C}{\mathcal N}\left( {0,\sigma _{c,u}^2} \right)$ denotes the additive white Gaussian noise (AWGN) at user $u$ with variance ${\sigma _{c,u}^2}$.
${\bf h}_u$ represents the channel form the DFRC BS to user $u$.

According to \eqref{eq:2}, the achievable rate of user $u$ is given by
\begin{equation}
	{\text{Rate}}_u \left( {\bf F}_{\rm A} , {\bf F}_{\rm D} \right) = \log \left( {1 + {\text{SIN}}{{\text{R}}_u}} \left( {\bf F}_{\rm A} , {\bf F}_{\rm D} \right) \right) ,
\end{equation}
where ${\text{SINR}}_u \left( {\bf F}_{\rm A} , {\bf F}_{\rm D} \right)$ is the signal-to-interference-plus-noise ratio (SINR) of the user $u$, which can be written as
\begin{equation}
	{\text{SIN}}{{\text{R}}_u} \left( {\bf F}_{\rm A} , {\bf F}_{\rm D} \right) = \frac{{{{\left| {{\bf{h}}_u^H{{\bf{F}}_{\text{A}}}{{\bf{f}}_{{\text{D}},u}}} \right|}^2}}}{{{{\sum\limits_{v \in {\mathcal U},v \ne u} {\left| {{\bf{h}}_u^H{{\bf{F}}_{\text{A}}}{{\bf{f}}_{{\text{D}},v}}} \right|} }^2} + \sigma _{c,u}^2}}	.
\end{equation}

\vspace{-1em}
\subsection{Metric for Radar Performance}
In addition to sending communication symbols to multiple users, the transmitting waveform simultaneously performs the radar sensing task.
It is assumed that the vehicle target of interests locates with angle $\theta_{\rm T}$ and the stationary clutters sources (such as trees, buildings, etc.) locate in angle $\theta_q, q\in \mathcal{Q}=\left\{1 , \cdots , Q\right\}$.
Based on the above assumption and the transmitting model \eqref{eq:1}, we collect the received signals during a radar pulse.
The received signals of $L$ time slots can be expressed as\footnote{We assume the vehicle target is slowly-moving or stay still. Hence, the Doppler is assumed to zero without loss of generality.}
\begin{equation}
	{{\bf{Y}}_r} = {\alpha_{\text{T}}}{\bf{A}}\left[ {{\theta _{\text{T}}}} \right]{{\bf{F}}_{\text{A}}}{{\bf{F}}_{\text{D}}}{\bf{S}} + \sum\limits_{q \in {\mathcal Q}} {{\alpha _q}{\bf{A}}\left[ {{\theta _q}} \right]{{\bf{F}}_{\text{A}}}{{\bf{F}}_{\text{D}}}{\bf{S}}}  + {{\bf{N}}_r},
	\label{eq:5}
\end{equation}
where ${{\bf{Y}}_r} = \left[ {{{\bf{y}}_r}\left[ 1 \right], \cdots ,{{\bf{y}}_r}\left[ L \right]} \right]$ , ${\bf{S}} = \left[ {{\bf{s}}\left[ 1 \right], \cdots ,{\bf{s}}\left[ L \right]} \right]$ and ${{\bf{N}}_r} = \left[ {{{\bf{n}}_r}\left[ 1 \right], \cdots ,{{\bf{n}}_r}\left[ L \right]} \right]$, the scalar $\alpha_{\text{T}}$ and $\alpha_{q}$ denote the target and clutter radar cross section (RCS) with 
$\alpha_{\text{T}} \sim \mathcal{CN}(0,\varsigma _\text{T})$ and $\alpha_q \sim \mathcal{CN}(0,\varsigma_q),\forall q \in \mathcal{Q}$.
The vector ${{\bf{n}}_r}\left[ l \right]$ is AWGN and ${{\bf{n}}_r}\left[ l \right] \sim \left( {{{\bf{0}}_{{N_{\text{R}}}}},\sigma _r^2{{\bf{I}}_{{N_{\text{R}}}}}} \right)$.
${\bf{A}}\left[ \theta  \right] = {{\bf{a}}_{\text{R}}}\left[ \theta  \right]{\bf{a}}_{\text{T}}^T\left[ \theta  \right]$ stands for the effective radar channel.
Particularly, for the half-wavelength spaced ULA, we have
${\bf{a}}\left[ \varphi  \right] = \sqrt {\frac{1}{N}} {\left[ {1,{e^{-\jmath 2\pi \sin \left[ \varphi  \right]}}, \cdots ,{e^{-\jmath 2\pi \left( {N - 1} \right) \sin \left[ \varphi  \right]}}} \right]^T}$.
After vectorizing DRFC BS received signals, the detection of the target at the cell under test (CUT) can be cast as the following binary hypothesis testing problem: 
\begin{equation}
	\left\{ 
	\begin{array}{ll}
		{\mathcal H}_1 : & {{\bf{y}}_{r}} = {{\bf{y}}_{\text{T}}} + {{\bf{y}}_{\text{C}}} + {{\bf{y}}_{\text{N}}} \\
		{\mathcal H}_0 : & {{\bf{y}}_{r}} = {{\bf{y}}_{\text{C}}} + {{\bf{y}}_{\text{N}}} \\
	\end{array}
	\right. ,
\end{equation}
where  ${{\bf{y}}_{\text{N}}} = {\text{Vec}}\left[ {{{\bf{N}}_r}} \right]$, ${{\bf{y}}_{\text{C}}} = \sum_{q \in {\mathcal Q}} {{\alpha _q}\left( {{\bf{I}}_L \otimes {\bf{A}}\left[ {{\theta _q}} \right]{{\bf{F}}_{\text{A}}}{{\bf{F}}_{\text{D}}}} \right){\bf{s}}} $ and ${{\bf{y}}_{\text{T}}} = {\alpha_{\text{T}}}\left( {{\bf{I}}_L \otimes {\bf{A}}\left[ {{\theta_{\text{T}}}} \right]{{\bf{F}}_{\text{A}}}{{\bf{F}}_{\text{D}}}} \right){\bf{s}}$ with ${\bf{s}} = {\text{Vec}}\left[ {\bf{S}} \right]$.

In this paper, we aim to improve the target detection performance in the presence of signal-dependent clutters.
Thus, we propose to apply RMI \cite{tang2010mimo} as the radar performance metric to evaluate how much useful information the DFRC BS can obtain from the received echoes.
The RMI is defined as \cite{tang2010mimo}
\begin{equation}
	\begin{aligned}
		{\text{RMI}} & \left( {{{\bf{y}}_r};{\bf{A}}\left[ {{\theta _T}} \right]|{{\bf{F}}_{\text{A}}},{{\bf{F}}_{\text{D}}}} \right)  = \frac{1}{{{N_{\text{R}}}L}}\left\{ {h\left( {{{\bf{y}}_r}|{{\bf{F}}_{\text{A}}},{{\bf{F}}_{\text{D}}}} \right) - h\left( {{{\bf{y}}_{{\text{CN}}}}} \right)} \right\} \\
		& = \log \left( {\det \left| {{\bf{I}}_{{N_{\text{R}}}}} + {\varsigma _{\text{T}}}{\bf{A}}\left[ {{\theta_{\text{T}}}} \right]{{\bf{F}}_{\text{A}}}{{\bf{F}}_{\text{D}}} {\bf{F}}_{\text{D}}^H{\bf{F}}_{\text{A}}^H{{\bf{A}}^H}\left[ {{\theta_{\text{T}}}}  \right]  {\bf{R}}_{{\text{CN}}}^{ - 1} \right|} \right) ,
	\end{aligned}
	\notag
\end{equation}
where ${{\bf{y}}_{{\text{CN}}}} = {{\bf{y}}_{\text{C}}} + {{\bf{y}}_{\text{N}}}$ and ${{\bf{R}}_{\rm CN}}$ is the effective covariance matrix of ${{\bf{y}}_{\rm{C}}}$ and ${{\bf{y}}_{\rm{N}}}$, which is defined as
\begin{equation}
	{{\bf{R}}_{{\text{CN}}}} = \sum\limits_{q \in {\mathcal Q}} {{\varsigma _q}{\bf{A}}\left[ {{\theta _q}} \right]{{\bf{F}}_{\text{A}}}{{\bf{F}}_{\text{D}}}{\bf{F}}_{\text{D}}^H{\bf{F}}_{\text{A}}^H{{\bf{A}}^H}\left[ {{\theta _q}} \right]}  + \sigma _r^2{{\bf{I}}_{{N_{\text{R}}}}} .
\end{equation}

\vspace{-0.5em}
\subsection{Problem Formulation}
To improve the radar detection performance while guaranteeing communication QoS requirements, we jointly design the transmit HBF with the proposed novel dynamic ABF architecture.
Specifically, we aim to maximize the RMI while ensuring 
\textit{i)} that the average achievable rate of each user is above a pre-defined threshold $\Gamma_u$\footnote{The QoS threshold $\Gamma_u$ should be chosen below the communication upper-bound \cite{Liu2022Integrated,zhang2021overview,zhou2022integrated} while meeting practical QoS requirements.}, 
\textit{ii)} that the transmitting waveform does not exceed the maximum allowable power $P$ and, 
\textit{iii)} that the ABF obeys the dynamic mapping criterion.
Therefore, the optimization problem is formulated as
\begin{equation}
	\begin{aligned}
		\mathop {\text{maximize}}\limits_{{{\bf{F}}_{\text{A}}} , {{\bf{F}}_{\text{D}}}} \quad & {\text{RMI}} \left( {\bf F}_{\rm A} , {\bf F}_{\rm D} \right) \\
		{\text{subject to}} \quad & {\text C1:} {\text{ Rate}}_u \left( {\bf F}_{\rm A} , {\bf F}_{\rm D} \right) \ge \Gamma_u , \forall u  \\
		& {\text C2:} \left\| {{{\bf{F}}_{\text{A}}}{{\bf{F}}_{\text{D}}}} \right\|_F^2 \le {{P}} , \; \\
		& {\text C3:} {{\bf{F}}_{\text{A}}} \in {\mathcal S} .
	\end{aligned}
	\label{eq:9}%
\end{equation}
Problem \eqref{eq:9} is a high-dimensional and non-convex optimization problem, which cannot be directly solved by existing algorithms.
In particular, the non-convexity stems from the highly coupled optimization variables, the log-fractional expression in both the objective function and constraint C1, and the dynamic mapping constraint C3 of the ABF matrix.
Therefore, a new algorithm is developed in the following section to solve this challenging problem. 

\vspace{-0.5em}
\section{Dynamic HBF Design for DFRC BS}

In this section, we transform the original problem into a more tractable form, propose an efficient algorithm to design dynamic HBF, and summarize the proposed algorithm.

\vspace{-0.5em}
\subsection{Problem Transformation}\label{sec:3-1}

First, we propose to simplify the complicated constraint C1 by exploiting the relationship between rate and mean-square error (MSE) \cite{shi2011iteratively} as mentioned in the following theorem.
\vspace{-0.5em}
\begin{theorem}
	By adopting the weighted minimum MSE (WMMSE) framework, the constraint {\rm{C1}} is equivalent to
	\begin{equation}
		{\rm{C}}1 \Leftrightarrow {{\dot{\rm{C}}1}: } - {w_u}{\mathcal E}\left( {{{\bf{F}}_{\rm{A}}},{{\bf{F}}_{\rm{D}}}} \right) + \log {w_u} + 1 \ge {\Gamma _u} ,
	\end{equation}
	where ${\mathcal E}\left( {{{\bf{F}}_{\text{A}}},{{\bf{F}}_{\text{D}}}} ,  \delta_u \right) = {\mathbb{E}}\left[ {{{\left| {{s_u}\left[ l \right] - {\delta _u}{y_u}\left[ l \right]} \right|}^2}} \right]$, $\delta_u$ and $w_u$ represent MSE, receive decoding coefficient and the weight for user $u$  \cite{shi2011iteratively}, respectively.
\end{theorem}
\vspace{-0.5em}
By introducing variables $w_u$ and $\delta_u$, $\forall u\in\mathcal{U}$ to the original constraint C1,  the new constraint ${\dot{\text{C}}1}$ is more tractable.
However, the highly coupling relationship between ${\bf F}_{\rm A}$  and ${\bf F}_{\rm D}$ still challenges the algorithm development.
Therefore, we introduce another auxiliary variable ${\bf T}$ and add the equality constraint $\mathbf{T} = \mathbf{F}_\mathrm{A}\mathbf{F}_\mathrm{D}$.
Then, problem \eqref{eq:9} is converted into
\begin{equation}
	\begin{aligned}
		\mathop {\text{maximize}}\limits_{ {\bf{T}} , {{\bf{F}}_{\text{A}}} , {{\bf{F}}_{\text{D}}} , \left\{ \delta_u \right\} , \left\{ w_u \right\} } \;\; &  {\overline{\text{RMI}}} \left( {\bf T} \right) \\
		{\text{subject to}} \;\; {\overline{\text C1}}: & - {w_u}{\mathcal E}\left( {\bf{T}} ,  \delta_u  \right) + \log {w_u} + 1 \ge {\Gamma _u} , \forall u , \\
		{\overline {\text C2}}: &  \left\| {{{\bf{T}}}} \right\|_F^2 \le {{P}} , \;\\
		{\text C3}:  & {{\bf{F}}_{\text{A}}} \in {\mathcal S} \\
		{\text C4:} & {{\bf{T}}} = {{\bf{F}}_{\text{A}}} {{\bf{F}}_{\text{D}}}  ,
	\end{aligned}
	\label{eq:10}
\end{equation}
where $\overline{{\text C1}}$ is obtained by replacing ${\bf F}_{\rm A}{\bf F}_{\rm D}$ in ${\dot{\text{C}}1}$ with $\bf T$, and  the objective function is reformulated as
\begin{equation}
	{\overline{\text{RMI}}}\left( {\bf{T}} \right) 
	= \log  {\det \left| {{{\bf{I}}_{{N_{\text{R}}}}} + \frac{{{\varsigma _{\text{T}}}{\bf{A}}\left[ {{\theta _{\text{T}}}} \right]{\bf{T}}{{\bf{T}}^H}{{\bf{A}}^H}\left[ {{\theta _{\text{T}}}} \right]}}{{\sum\limits_{q \in {\mathcal Q}} {{\varsigma _q}{\bf{A}}\left[ {{\theta _q}} \right]{\bf{T}}{{\bf{T}}^H}{{\bf{A}}^H}\left[ {{\theta _q}} \right]}  + \sigma _r^2{{\bf{I}}_{{N_{\text{R}}}}}}}} \right|}
	\nonumber
\end{equation}

Introducing the Lagrangian dual variable ${\bm{\Lambda }} \in {\mathbb C}^{N_{\rm T}\times N_{\rm U}}$ and the penalty parameter $\rho \ge 0$ for the additional equality constraint C4, we formulate the following augmented Lagrangian (AL) problem of \eqref{eq:10} as
\begin{equation}
	\begin{aligned}
		\mathop {\text{minimize}}\limits_{ {\bf{T}} , {{\bf{F}}_{\text{A}}} , {{\bf{F}}_{\text{D}}} , \left\{ \delta_u \right\} , \left\{ w_u \right\} } \;\; &  {\mathcal L}\left( {{\bf{T}},{{\bf{F}}_{\text{A}}},{{\bf{F}}_{\text{D}}}} \right) \\
		{\text{subject to}} \qquad  &  {\overline{\text C1}} , {\overline {\text C2}} , {\text C3} ,  \; {\text{and}} \; {\text C4},
	\end{aligned}
	\label{eq:11}
\end{equation}
where ${\mathcal L}\left( {{\bf{T}},{{\bf{F}}_{\text{A}}},{{\bf{F}}_{\text{D}}}} \right) =  - \overline {{\text{  RMI}}} \left( {\bf{T}} \right) + \frac{\rho }{2}\left\| {{\bf{T}} - {{\bf{F}}_{\text{A}}}{{\bf{F}}_{\text{D}}}} \right\|_F^2 + \Re \left( {{\text{Tr}}\left[ {{{ \bm\Lambda} ^H}\left( {{\bf{T}} - {{\bf{F}}_{\text{A}}}{{\bf{F}}_{\text{D}}}} \right)} \right]} \right)$.

Now the original problem \eqref{eq:9} is transformed into the AL problem \eqref{eq:11} with multiple variables, whose constraints are separated from each other. 
The AO algorithm is invoked to sequentially address each variable by fixing other variables, which leads to the following subproblems.

\vspace{-.5em}
\subsection{Solution to Subproblems}
\subsubsection{Update ${\bf T}$}
With the fixed ${\bf F}_{\rm A}$, ${\bf F}_{\rm D}$, $\left\{ \delta_u \right\}_{\forall u}$, $\left\{w_u\right\}_{\forall u}$, and ${\bm \Lambda}$, the optimization problem for updating ${\bf T}$ is formulated as
\begin{subequations}
	\begin{align}
		\mathop {\text{minimize}}\limits_{ {\bf{T}} } \quad & - {\overline{\text{RMI}}} \left( {\bf T} \right) + \frac{\rho}{2} \left\| {{\bf{T}} - {{\bf{F}}_{\text{A}}}{{\bf{F}}_{\text{D}}}} \right\|_F^2 \notag \\ 
		& \quad + \Re \left( {{\text{Tr}}\left[ {{{\bf{\Lambda }}^H}\left( {{\bf{T}} - {{\bf{F}}_{\text{A}}}{{\bf{F}}_{\text{D}}}} \right)} \right]} \right) \label{eq:12a}\\
		{\text{subject to}} \quad & {\overline{\text C1}}, \; {\rm and} \; {\overline {\text C2}} .
	\end{align}
	\label{eq:12}%
\end{subequations}
To tackle the complicated objective function from problem \eqref{eq:12}, we adopt the majorization-minimization (MM) \cite{sun2016majorization} method to simplify the objective by the following lemmas.
\vspace{-1.5em}
\begin{lemma}\label{lem:1}
	For any positive defined matrix ${\bf A} \in {\mathbb C}^{M\times N}$, and ${\bf B} \in {\mathbb C}^{M\times N}$, we have
	\begin{equation}
		\log \det \left| {{\bf{I}}_{N} + {{\bf{A}}^H}{{\left[ {{\bf{B}}{{\bf{B}}^H}} \right]}^{ - 1}}{\bf{A}}} \right| = \log \det \left| {{{\bf{E}}^H}{{\bf{C}}^{ - 1}}{\bf{E}}} \right| ,
	\end{equation}
	where ${\bf{C}} = \left[ \begin{array}{cc}
		{{{\bf{I}}_N}} & {{{\bf{A}}^H}} \\
		{\bf{A}} & { {{\bf{A}}^H}{\bf{A}} + {{\bf{B}}^H}{\bf{B}} }
	\end{array} \right]$ and ${\bf{E}} = \left[ {{{\bf{I}}_N};{\bf{0}}_{M\times N}} \right]^T$.
\end{lemma}
\vspace{-0.5em}
\begin{IEEEproof}
	The proof is a straightforward application of the inversion lemma of a partitioned matrix \cite[Eq. (1.7.2)]{zhang2017matrix}.
\end{IEEEproof}

Based on \textbf{Lemma \ref{lem:1}}, the ${\overline{\text{RMI}}} \left( {\bf T} \right)$ in the objective function \eqref{eq:12a} can be reformulated as
\begin{equation}
	\overline {{\text{RMI}}} \left( {\bf{T}} \right) = \log \det \left| {{{\bf{E}}^H}{{\left[ {{\bf{\Xi }}\left( {\bf{T}} \right)} \right]}^{ - 1}}{\bf{E}}} \right|,\label{eq:14}
\end{equation}
where the fresh notations are defined as ${\bf{E}} = {\left[ {{{\bf{I}}_{{N_{\text{U}}}}},{{\bf{0}}_{{N_{\text{U}}} \times {N_{\text{R}}}}}} \right]^T}$ and ${\bm{\Xi}}\left( {\bf{T}} \right) = \left[ \begin{array}{cc}
	{{{\bf{I}}_{{N_{\text{U}}}}}} & {\sqrt {{\varsigma _{\text{T}}}} {{\bf{T}}^H}{{\bf{A}}^H}\left[ {{\theta _{\text{T}}}} \right]} \\
	{\sqrt {{\varsigma _{\text{T}}}} {\bf{A}}\left[ {{\theta _{\text{T}}}} \right]{\bf{T}}} & {{{\bf{\hat R}}}_{tc}} + \sigma _r^2{{\bf{I}}_{{N_{\text{R}}}}}
\end{array} \right]$ with ${{{\bf{\hat R}}}_{tc}} = \sum_{q \in \left\{ {{\text{T}},{\mathcal Q}} \right\}} {{\varsigma _q}{\bf{A}}\left[ {{\theta _q}} \right]{\bf{T}}{{\bf{T}}^H}{{\bf{A}}^H}\left[ {{\theta _q}} \right]} $.
We further introduce the following lemma to find the lower-bound of objective \eqref{eq:14}.

\vspace{-0.5em}
\begin{lemma}\label{lem:2}
	For any positive defined Hermitian matrix ${\bf C}$, $\log \det | {{{\bf{E}}^H}{{\bf{C}}^{ - 1}}{\bf{E}}} |$ is a convex function with respect to ${\bf C}$, when $\bf E$ is full column rank matrix.
	A minorizer of $\log \det | {{{\bf{E}}^H}{{\bf{C}}^{ - 1}}{\bf{E}}} |$ is defined as
	\begin{equation}
		\begin{aligned}
			\log \det  \left| {{{\bf{E}}^H}{{\bf{C}}^{ - 1}}{\bf{E}}} \right| \ge  
			\log \det \left| {{{\bf{E}}^H}{\bf{C}}_k^{ - 1}{\bf{E}}} \right| 
			- {\rm{Tr}}\left[ {{\bf{G}}\left( {{\bf{C}} - {\bf{C}}_k} \right)} \right] ,
		\end{aligned}
		\nonumber
	\end{equation}
	where ${\bf C}_k$ is the obtained ${\bf C}$ at the $k^{\rm th}$ iteration, and ${\bf{G}} = {\bf{C}}_k^{ - 1}{\bf{E}}\left[ {{{\bf{E}}^H}{\bf{C}}_k^{ - 1}{\bf{E}}} \right]^{-1}{{\bf{E}}^H}{\bf{C}}_k^{ - 1}$.
\end{lemma}
\vspace{-0.5em}
\begin{IEEEproof}
	See \cite[Appendix A]{naghsh2017information}.
\end{IEEEproof}

Based on \textbf{Lemma \ref{lem:2}}, the following equality holds true
\begin{equation}
	\begin{aligned}
		&  \overline {{\text{RMI}}} \left( {\bf{T}} \right) = \log \det \left| {{{\bf{E}}^H}{{\left[ {{\bf{\Xi }}\left( {\bf{T}} \right)} \right]}^{ - 1}}{\bf{E}}} \right| \\
		& \ge \log \det \left| {{{\bf{E}}^H}{{\left[ {{\bf{\Xi }}\left( {{{\bf{T}}_k}} \right)} \right]}^{ - 1}}{\bf{E}}} \right| 
		- {\text{Tr}}\left[ {{\bf{G}}_k\left( {{\bf{\Xi }}\left( {\bf{T}} \right) - {\bf{\Xi }}\left( {{{\bf{T}}_k}} \right)} \right)} \right] \\
		& = {\text{Tr}}\left[ {\bf G}_k{{\bf{\Xi }}\left( {\bf{T}} \right)} \right] + {\rm const.}
	\end{aligned}
	\nonumber
\end{equation}
where ${{\bf{G}}_k} = {[ {{\bf{\Xi }}( {{{\bf{T}}_k}})} ]^{ - 1}}{\bf{E}}{[ {{{\bf{E}}^H}{{[ {{\bf{\Xi }}( {{{\bf{T}}_k}} )} ]}^{ - 1}}{\bf{E}}} ]^{ - 1}}{{\bf{E}}^H}{[ {{\bf{\Xi }}( {{{\bf{T}}_k}} )} ]^{ - 1}}$.

By defining ${\bf t} = {\rm Vec} \left[ {\bf T} \right]$ and blocking $\mathbf{G}_k$ as
${\bf G}_k = \left[\begin{array}{cc}
	{{\bf{G}}_k^{\left( {1,1} \right)}}   \;  {{\bf{G}}_k^{\left( {1,2} \right)}}  \\
	{{\bf{G}}_k^{\left( {2,1} \right)}}   \;  {{\bf{G}}_k^{\left( {2,2} \right)}}  
\end{array}
\right]
$, 
${\text{Tr}}\left[ {\bf G}_k {{\bf{\Xi }}\left( {\bf{T}} \right)} \right]$ is equivalently derived as
\begin{equation}
	\begin{aligned}
		{\text{Tr}} & \left[ {{{\bf{G}}_k}{\bf{\Xi }}\left( {\bf{T}} \right)} \right] = {\text{Tr}}\left[ {{\bf{G}}_k^{\left( {2,2} \right)} \left( {{{{\bf{\hat R}}}_{tc}}}   + \sigma _r^2{{\bf{I}}_{{N_{\text{R}}}}} \right) } \right] + {\text{Tr}}\left[ {{\bf{G}}_k^{\left( {1,1} \right)}} \right]   \\ 
		& + 2\Re \left\{ {\sqrt {{\varsigma _{\text{T}}}} {\text{Tr}}\left[ {{\bf{G}}_k^{\left( {1,2} \right)}{\bf{A}}\left[ {{\theta _{\text{T}}}} \right]{\bf{T}}} \right]} \right\} 
		=  {{\bf{t}}^H}{{{\bf{\hat Q}}}_k}{\bf{t}} - \Re \left( {{\bf{\hat q}}_k^H{\bf{t}}} \right)  ,
	\end{aligned}
	\nonumber
\end{equation}
where ${{{\bf{\hat q}}}_k} = {\text{Vec}}[ {2\sqrt {{\varsigma _{\text{T}}}} {{\bf{A}}^H}\left[ {{\theta _{\text{T}}}} \right]{( {{\bf{G}}_k^{\left( {1,2} \right)}} )^H} - \rho {{\bf{F}}_{\text{A}}}{{\bf{F}}_{\text{D}}} + {\bf{\Lambda }}} ]$ and ${{{\bf{\hat Q}}}_k} = [ {{{\bf{I}}_{{N_{\text{U}}}}} \otimes ( {\sum_{q \in \left\{ {{\text{T}},{\mathcal Q}} \right\}} {{\varsigma _q}{{\bf{A}}^H}\left[ {{\theta _q}} \right]{\bf{G}}_k^{\left( {2,2} \right)}{\bf{A}}\left[ {{\theta _q}} \right]}  + \rho {{\bf{I}}_{{N_{\text{T}}}}}} )} ]$.

Based on the above MM for objective \eqref{eq:14} and some algebraic manipulations for constraint $\overline {{\text{C1}}}$, the problem for updating ${\bf T}$ can be reformulated as follows
\begin{equation}
	\begin{aligned}
		\mathop {\text{minimize}}\limits_{ {\bf{t}} } \quad & {{\bf{t}}^H}{{{\bf{\hat Q}}}_k}{\bf{t}} - \Re \left( {{\bf{\hat q}}_k^H{\bf{t}}} \right)\\
		{\text{subject to}} \quad & {\widetilde{\text C1}}: {{\bf{t}}^H}{{\bf{M}}_u}{\bf{t}} - \Re \left( {{\bf{m}}_u^H{\bf{t}}} \right) + {\Gamma _u} \le 0 , \forall u  \\
		& {\overline {\text C2}}: {{\bf{t}}^H}{\bf{t}} \le {P} .
	\end{aligned}
	\label{eq:18}
\end{equation}
where 
${{\bf{M}}_u} = {w_u}\delta _u^2\left[ {{\bf{I}} \otimes {{\bf{h}}_u}{\bf{h}}_u^H} \right]$, ${{\bf{m}}_u} = 2{w_u}\delta _u^*{\text{Vec}}\left[ {{{\bf{h}}_u}{\bf{e}}_u^T} \right]$, and
${\bf{e}}_u$ is $u^{\rm th}$ column of identity matrix ${\bf I}_{N_{\rm U}}$.
Problem \eqref{eq:18} is a quadratic constraint quadratic programming (QCQP) problem, which can be efficiently solved by many existing approaches.

\subsubsection{Update ${\bf F}_{\rm A}$}\label{sec:3-3}

With the fixed ${\bf T}$, ${\bf F}_{\rm D}$, $\left\{ \delta_u \right\}_{\forall u}$, $\left\{w_u\right\}_{\forall u}$, and ${\bm \Lambda}$, the optimization problem for updating ${\bf F}_{\rm A}$ with dynamically sub-connected architecture is formulated as
\begin{equation}
	\begin{aligned}
		\mathop {\text{minimize}}\limits_{ {\bf{F}}_{\rm A} } \quad & \frac{\rho}{2} \left\| {{\bf{T}} - {{\bf{F}}_{\text{A}}}{{\bf{F}}_{\text{D}}}} \right\|_F^2 + \Re \left( {{\text{Tr}}\left[ {{{\bf{\Lambda }}^H}\left( {{\bf{T}} - {{\bf{F}}_{\text{A}}}{{\bf{F}}_{\text{D}}}} \right)} \right]} \right) \\
		{\text{subject to}} \quad & {{\text{C3-1:}}} \left\| {{{\bf{F}}_{\text{A}}}\left[ {i,:} \right]} \right\|_0 = 1 , \;
		{{\text{C}}} \text{3-2: } \left| {{{\bf{F}}_{\text{A}}}\left[ {i,j} \right]} \right| \le 2  .
	\end{aligned}
	\label{eq:19}
\end{equation}
Constraint C3-1 indicates that there is only one non-zero element in each row of ${\bf F}_{\rm A}$, which motivates us to solve problem \eqref{eq:19} row-by-row.
Specifically, the problem for determining element in $i^{\rm th}$ row and $j^{\rm th}$ column is given by
\begin{equation}
	\begin{aligned}
		\mathop {\text{minimize}}\limits_{ {\bf{F}}_{\rm A} \left[ i , j \right] } \;\; & {\mathcal K}\left( {i,j} \right) = \left\| {{{\bf{F}}_{\text{D}}}\left[ {j,:} \right]} \right\|_F^2{\left| {{{\bf{F}}_{\text{A}}}\left[ {i,j} \right]} \right|^2}  \\
		& - 2\Re \left\{ {{{\bf{F}}_{\text{A}}}\left[ {i,j} \right]{{\bf{F}}_{\text{D}}}\left[ {j,:} \right]{{{\bf{\tilde T}}}^H}\left[ {i,:} \right]} \right\} + {\text{const.}} \\
		{\text{subject to}} \;\; & \left| {{{\bf{F}}_{\text{A}}}\left[ {i,j} \right]} \right| \le 2.
	\end{aligned}
\end{equation}
whose closed-form solution can be obtained as
\begin{equation}
	{{{\bf{F}}_{\text{A}}}\left[ {i,j} \right]} = \left\{  
	\begin{array}{ll}
		{A_{i,j}}{e^{\jmath {\phi _{i,j}}}}  , &  {A_{i,j}} \le 2 , \\
		2{e^{\jmath {\phi _{i,j}}}} , & \text{otherwise.}
	\end{array}
	\right. ,
\end{equation}
where ${\phi _{i,j}} = \angle \{ {{\bf{\tilde T}}[ { i,:}]{\bf{F}}_{\text{D}}^H[ { j,:}]}\}$, ${A_{i,j}} = \frac{{\left| {{{\bf{F}}_{\text{D}}}\left[ { j,:} \right]{{{\bf{\tilde T}}}^H}\left[ { i,:} \right]} \right|}}{{\left\| {{{\bf{F}}_{\text{D}}}\left[ { j,:} \right]} \right\|_F^2}}$, and ${\bf{\tilde T}} = {\bf{T}} + {{\bm{\Lambda }}}/{\rho }$.
After solving the problem for all elements in $i^{\rm th}$ row, we select $j_\star^{\rm th}$ element which satisfies $j_\star \leftarrow \mathop {\min }_j {\mathcal K}\left( {i,j} \right)$ as the non-zero element in $i^{\rm th}$ row, i.e., ${{\bf{F}}_{\text{A}}}\left[ {i,j} \right] = {{\bf{F}}_{\text{A}}}\left[ {i,{j_ \star }} \right],{{\bf{F}}_{\text{A}}}\left[ {i,j} \right] = 0 \;\forall j \ne {j_ \star }$.
Finally, after obtaining the optimal solution of ABF, we can arrange the value of each pair DPS as
\begin{equation}
	\phi _{i,j}^1 =  {\phi _{i,j}} + \arccos \left( {{A_{i,j}}/2} \right),	
	\phi _{i,j}^2 = {\phi _{i,j}} - \arccos \left( {{A_{i,j}}/2} \right).
	\nonumber
\end{equation}

\subsubsection{Update ${\bf F}_{\rm D}$}
With the fixed ${\bf T}$, ${\bf F}_{\rm A}$, $\left\{ \delta_u \right\}_{\forall u}$, $\left\{w_u\right\}_{\forall u}$, and ${\bm \Lambda}$, the optimization problem for updating ${\bf F}_{\rm D}$ is formulated as
\begin{equation}
	\mathop {\text{minimize}}\limits_{ {\bf{F}}_{\rm D} } \; \frac{\rho}{2} \left\| {{\bf{T}} - {{\bf{F}}_{\text{A}}}{{\bf{F}}_{\text{D}}}} \right\|_F^2 + \Re \left( {{\text{Tr}}\left[ {{{\bf{\Lambda }}^H}\left( {{\bf{T}} - {{\bf{F}}_{\text{A}}}{{\bf{F}}_{\text{D}}}} \right)} \right]} \right) ,
\end{equation}
whose optimal solution is obtained via first order derivative as
\begin{equation}
	{{\bf{F}}_{\text{D}}} = {\left[ {{\bf{F}}_{\text{A}}^H{{\bf{F}}_{\text{A}}}} \right]^{ - 1}}{\bf{F}}_{\text{A}}^H\left[ {{\bf{T}} + {{\bf{\Lambda }}}/{\rho }} \right] .
	\label{eq:21}
\end{equation}

\subsubsection{Update Auxiliary Variables}
Finally, with fixed ${\bf T}$, ${\bf F}_{\rm A}$, ${\bf F}_{\rm D}$, we need to update ${\left\{ w_u \right\}}_{\forall u} , {\left\{ \delta_u \right\}}_{\forall u}$ whose optimal solutions can be derived by first order derivative \cite{shi2011iteratively} as 
\begin{subequations}
	\begin{align}
		\delta _u & = {\Big( {\sum\limits_{v \in {\mathcal U}} {{{\left| {{\bf{h}}_u^H{{\bf{t}}_v}} \right|}^2}}  + \sigma _c^2} \Big)^{ - 1}}{\bf{t}}_u^H{{\bf{h}}_u}, \forall u  ,\label{eq:22a} \\
		w_u & = {\Big( {\sum\limits_{v \in {\mathcal U},v \ne u} {{{\left| {{\bf{h}}_u^H{{\bf{t}}_v}} \right|}^2}}  + \sigma _{c,u}^2} \Big)^{ - 1}}{\left| {{\bf{h}}_u^H{{\bf{t}}_u}} \right|^2} + 1, \forall u  . \label{eq:22b}
	\end{align}
	\label{eq:22}
\end{subequations}

\setlength{\textfloatsep}{0pt}
\begin{algorithm}[!tb]
	\caption{Dynamic HBF Design for DFRC BS}
	\label{alg:1}
	\LinesNumbered
	\KwIn{${\bf T}^{\left[ 0 \right]}$, ${\bf F}_{\text A}^{\left[ 0 \right]}$, ${\bf F}_{\text D}^{\left[ 0 \right]}$, $\{\delta_u^{\left[0\right]}\}$, $\{w_u^{\left[0\right]}\}$, $k = 0$ and \hspace{-0.2em}$\rho \ge 0$.}
	\While{no convegence}{
		$k = k + 1$\;
		
		Update ${\bf T}$ by QCQP problem \eqref{eq:18}  \;
		
		Update ABF ${\bf F}_{\rm A}$ by solving \eqref{eq:19} \;
		
		Update DBF ${\bf F}_{\rm D}$ by \eqref{eq:21} \;

		Update $\left\{\delta_u\right\}_{\forall u}$ and ${\left\{w_u\right\}}_{\forall u}$ as in \eqref{eq:22}  \;
		
		${\bf{\Lambda }} = {\bf{\Lambda }} + \rho \left[ {{\bf{T}} - {{\bf{F}}_{\text{A}}}{{\bf{F}}_{\text{D}}}} \right]$\;
	}
\end{algorithm}

\vspace{-2em}
\subsection{Summary}

Based on the above derivations, the proposed algorithm is summarized as \textbf{Algorithm \ref{alg:1}}. 
The main computational complexity of the overall algorithm is dominated by step 3-5 of \textbf{Algorithm \ref{alg:1}}.
Updating ${\bf T}$ is a QCQP problem, whose computational complexity is upper bound by ${\mathcal O}\left( {N_{\text{T}}^{3}N_{\text{U}}^{3}} \right)$. 
Updating ${\bf F}_{\rm A}$ has the complexity of ${\mathcal O}\left( {N_{\text{T}}^2{N_{{\text{RF}}}^2}N_{\text U}} \right)$.
Updating ${\bf F}_{\rm D}$ with close form solution requires complexity of ${\mathcal O}\left( { {N_{\text{T}}}N_{{\text{RF}}}^2 + N_{\text{T}}^2{N_{{\text{RF}}}}} \right)$.
Therefore, the overall complexities of the proposed algorithm is of ${\mathcal O}\left( {{I_0}\left( {N_{\text{T}}^3N_{\text{U}}^3 + {N_{\text{T}}^2{N_{{\text{RF}}}^2}N_{\text U}}} \right)} \right)$, where $I_0$ is number of outer iteration.

\begin{table*}[!tb]
	\centering
	\caption{Comparison for HBF Design for DFRC with Different ABF Architecture.}
	\vspace{-0.5em}
	\begin{tabular}{|c|c|c|c|}
		\hline
		Architecture  &                   
        No. of PSs    &        
        Computational Complexity   &
        Power Consumption $P_{\rm TOL}$
        \\ \hline \hline
        
		FC 					& 
        $N_{\rm T}N_{\rm RF}$ & 
        ${\mathcal O}\left( {{I_0}\left( {N_{\text{T}}^3N_{\text{U}}^3 + {I_{{\text{MO}}}}{N_{\text{T}}^2}{N_{{\text{RF}}}}{N_{\text{U}}}} \right)} \right)$ &
        $P + N_{\text{RF}}P_{\text{RF}} + N_{\rm T}N_{\text{RF}}P_{\text{PS}} + P_{\text{BB}}$
        \\ \hline
        
		Fix-SPS 	
        & $N_{\rm T}$           
        & ${\mathcal O}\left( {{I_0}N_{\text{T}}^{3}N_{\text{U}}^{3}} \right)$    &
        $P + N_{\text{RF}}P_{\text{RF}} + N_{\rm T}P_{\text{PS}} + P_{\text{BB}}$
        \\ \hline
        
		Fix-DPS   	& 
        $2N_{\rm T}$          & 
        ${\mathcal O}\left( {{I_0}N_{\text{T}}^{3}N_{\text{U}}^{3}} \right)$   &
        $P + N_{\text{RF}}P_{\text{RF}} + 2N_{\rm T}P_{\text{PS}} + P_{\text{BB}}$
        \\ \hline
        
		Dym-SPS						& 
        $N_{\rm T}$           & 
        ${\mathcal O}\left( {{I_0}\left( {N_{\text{T}}^3N_{\text{U}}^3 + {N_{\text{T}}^2{N_{{\text{RF}}}^2}N_{\text U}} } \right)} \right)$           &
        $P + N_{\text{RF}}P_{\text{RF}} + N_{\rm T}P_{\text{PS}} + P_{\text{BB}} + N_{\rm T}P_{\text{SW}}$
        \\ \hline
        
		Dym-DPS					& 
        $2N_{\rm T}$          & 
        ${\mathcal O}\left( {{I_0}\left( {N_{\text{T}}^3N_{\text{U}}^3 + {N_{\text{T}}^2{N_{{\text{RF}}}^2}N_{\text U}} } \right)} \right)$           &
        $P + N_{\text{RF}}P_{\text{RF}} + 2N_{\rm T}P_{\text{PS}} + P_{\text{BB}} + N_{\rm T}P_{\text{SW}}$
        \\ \hline
	\end{tabular}
	\label{tab:1}
	\vspace{-0.5em}
\end{table*}

\vspace{-1em}
\section{Simulation Results}

\begin{figure*}[t]
	\centering
	\begin{minipage}{0.31\linewidth}
		\vspace{-0.2em}
		\centering
		\includegraphics[width=1\linewidth]{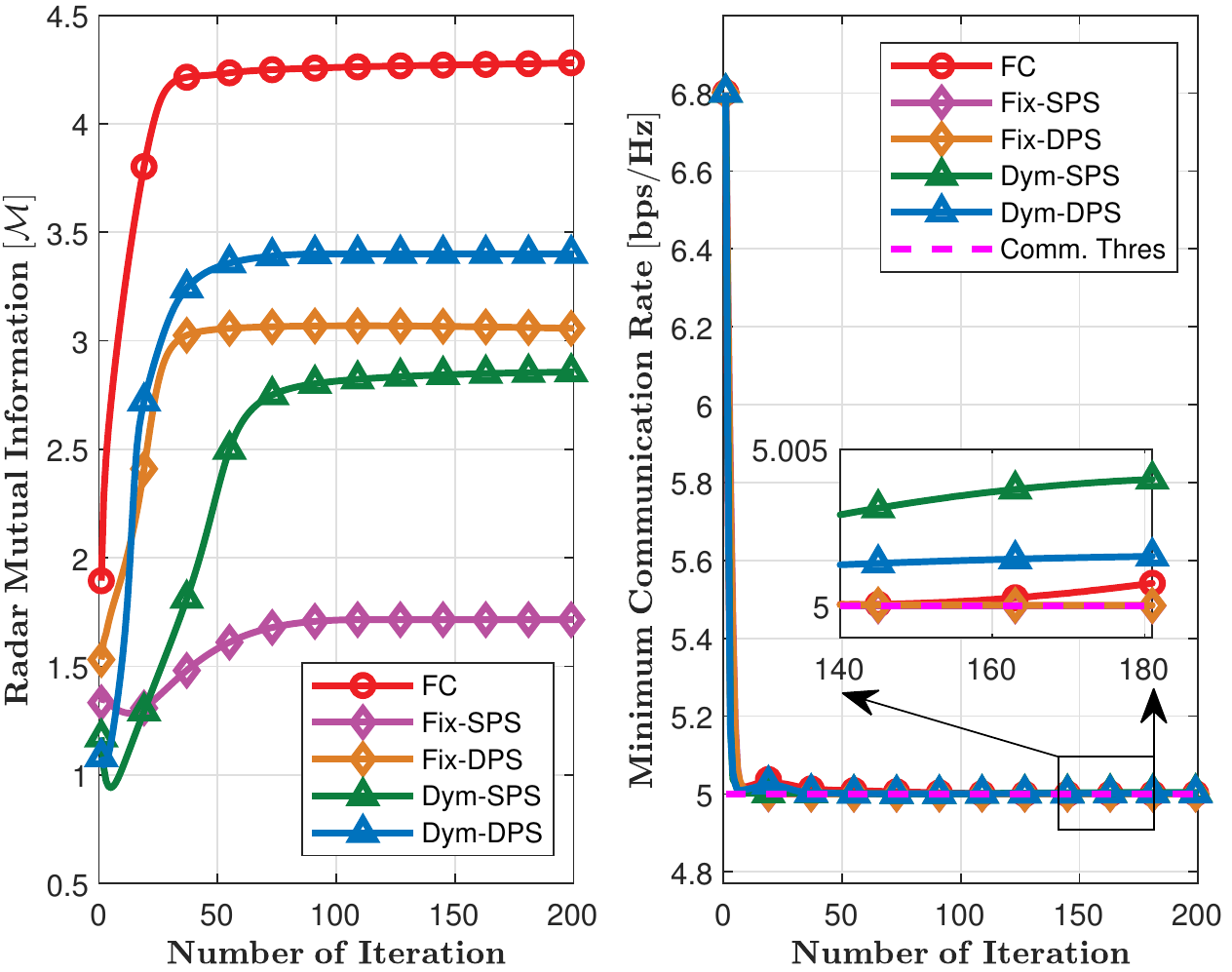}
		\vspace{-2em}
		\caption{Convergence comparisons with QoS threshold $\Gamma = 5$.}
		\label{fig:sim0}
	\end{minipage}
	\hspace{1em}
	\begin{minipage}{0.31\linewidth}
		\centering
		\includegraphics[width=1\linewidth]{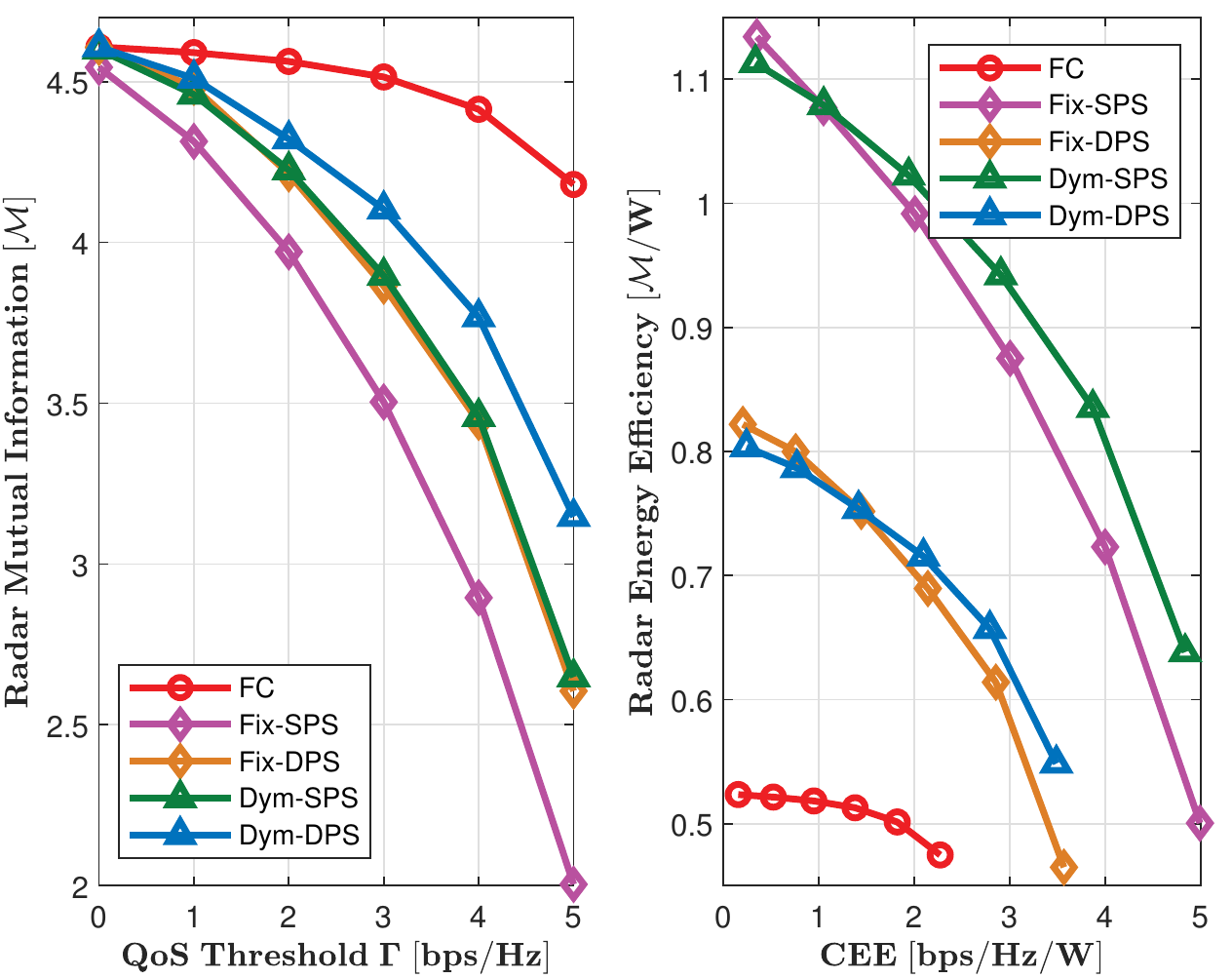}
		\vspace{-2em}
        \caption{Performance evaluation. Left: RMI vs. QoS threshold $\Gamma$. Right: REE vs. CEE.}
		\label{fig:sim1}
	\end{minipage}
	\hspace{1em}
	\begin{minipage}{0.31\linewidth}
		\centering
		\includegraphics[width=1\linewidth]{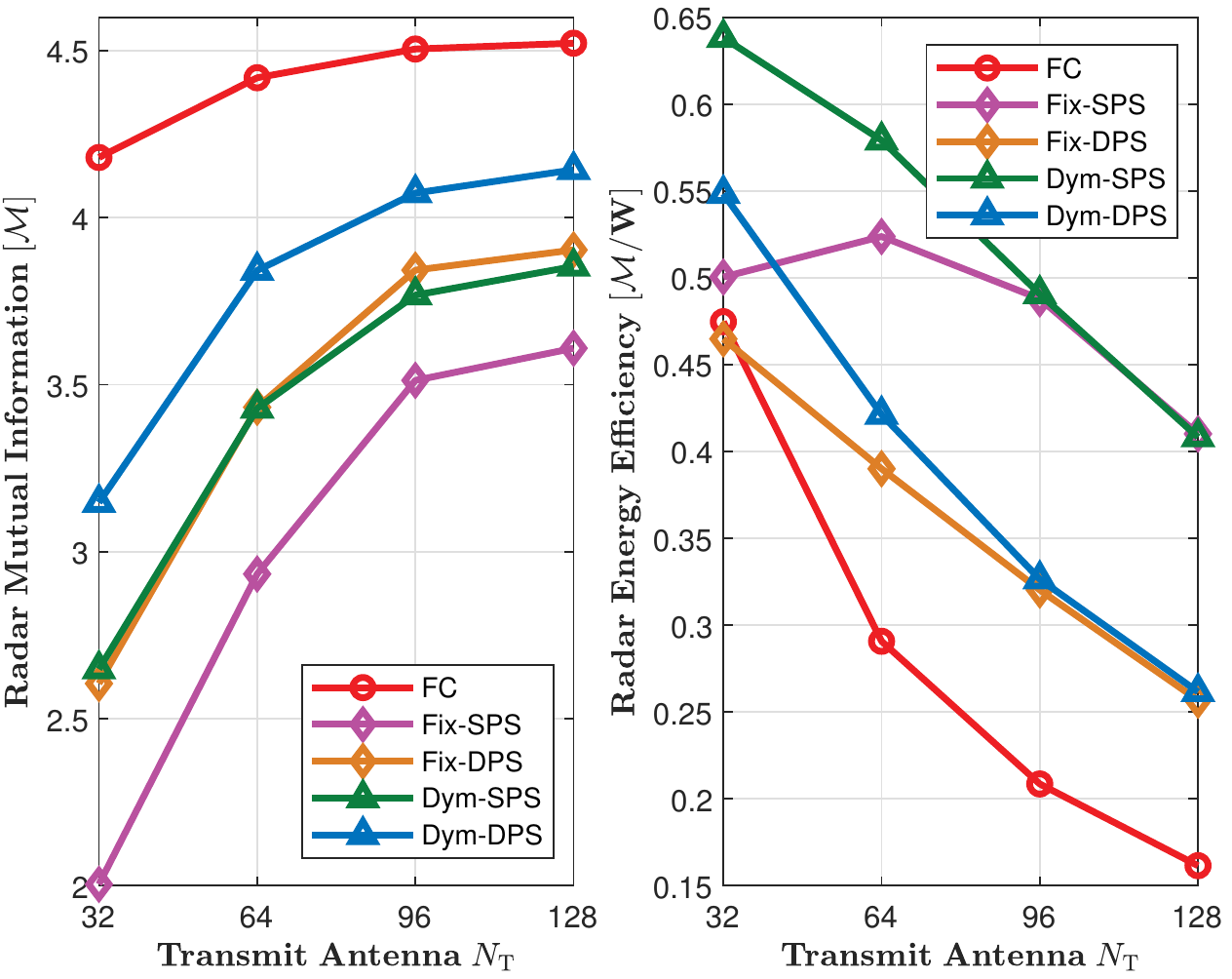}
		\vspace{-2em}
		\caption{Impact of $N_{\rm T}$ with QoS threshold $\Gamma = 5$. Left: RMI vs. $N_{\rm T}$, Right: REE vs. $N_{\rm T}$.}
		\label{fig:sim4}
	\end{minipage}
	\vspace{-0.5em}
\end{figure*}

The following settings are assumed throughout our simulations unless otherwise specified.
The DFRC BS equipped with ${N_{\rm T}} = 32$ transmit and ${N_{\rm R}} = 4$ receive antennas transmits $N_{\rm S}$ data streams to serve $N_{\rm U} = 4$ downlink users.
We assume the DFRC BS adopts $N_{\rm {RF}}$ RF chains, which satisfies $N_{\rm {RF}} = N_{\rm S} = N_{\rm U} = 4$.
The available transmit power is ${P} = 1 \text{W}$ and the number of time slot in considering frame is $L=8$.
The communication QoS threshold $\Gamma_u$ and noise power $\sigma_{c,u}^2$ at users are set same, i.e., $\Gamma = \Gamma_u$ and $\sigma_{c}^2 = \sigma_{c,u}^2$, which satisfy the $\text{SNR}_u = {{P}}/{\sigma_{c,u}^2} = \text{15dB}, \forall u$.
For the radar function, we assume the DFRC BS detects a target located in angle $\theta_{\rm T} = 0^\circ$, in presence of $Q = 3$ clutters, which are located in angles $\theta_1 = -50^\circ$, $\theta_2 = -10^\circ$, and $\theta_3 = 40^\circ$, respectively.
The RCS for the target and clutter sources are set as $\varsigma_{\rm T} = 20 \text{dB}$ and $\varsigma_q = 30 \text{dB},\forall q$, respectively.
The radar noise power $\sigma_r$ is set as $\sigma_r^2 = \text{0dB}$.
In the hardware power consumption model, $P_{\rm RF}$ and $P_{\rm BB}$ are the power consumed by each RF chain and baseband beamformer, $P_{\rm PS}$ and $P_{\rm SW}$ are the power consumption of a single PS and switch.
Particularly, we adopt the typical values $P_{\rm RF} = 300$mW, $P_{\rm BB} = 200$mW, $P_{\rm PS} = 50$mW, and $P_{\rm SW} = 5$mW \cite{li2020dynamic}.

To verify the superiority of the proposed dynamic HBF with DPS (Dym-DPS), the following benchmarks are included\footnote{To achieve fair comparisons, we fix the QoS requirements and compare the radar performance improvement of different architectures.}:
\textit{i}) fully-connected (FC) HBF \cite{qi2022hybrid}, where each RF chain is connected to all transmit antennas through SPS;
\textit{ii}) fixed sub-connected HBF with SPS (Fix-SPS) \cite{cheng2021hybrid}, where each RF chain is connected to a fixed subset of transmit antennas through SPS;
\textit{iii}) fixed sub-connected HBF with DPS (Fix-DPS) \cite{cheng2021double}, where each RF chain is connected to a fixed subset of transmit antennas through DPS;
\textit{iv}) dynamically sub-connected HBF with SPS (Dym-SPS) \cite{li2020dynamic}, where each RF chain is connected to a dynamic subset of transmit antennas through SPS.
The above benchmarks can be obtained by modifying the step for updating the ABF ${\bf F}_{\rm A}$ of our proposed algorithm.
For clarity, we summarize the number of PSs, optimization complexity, and power consumption for different architectures in \textbf{Table \ref{tab:1}}.

In Fig. \ref{fig:sim0}, we evaluate the convergence of the proposed algorithm for the considered novel dynamic HBF architecture with $\Gamma = 5$.
The left side of Fig. \ref{fig:sim0} \footnote{Since there is no universally agreed-upon unit for RMI, in this paper, we use $\mathcal{M}$ as RMI unit without loss of generality \cite{tang2010mimo}.} shows that the proposed algorithm converges to a stationary point for all cases.
The right side of Fig. \ref{fig:sim0} shows that the QoS requirement for different HBF architectures can always be guaranteed, which verifies the effectiveness of the proposed algorithm.
Furthermore, the proposed Dym-DPS architecture needs roughly 20 additional iterations to converge to a small objective value compared to the FC.
However, it significantly outperforms other considered sub-connected ABF architectures.

In the left side of Fig. \ref{fig:sim1}, we study the RMI versus the communication QoS threshold $\Gamma$ for different ABF architectures.
As expected, the system radar mutual information decreases monotonically with the communication QoS requirement.
This is because as $\Gamma$ increases, the DFRC BS system becomes less flexible in improving the radar performance to guarantee the QoS requirements.
Moreover, the proposed dynamically sub-connected HBF architecture always outperforms its competitors over the considered QoS range.
In the right side of Fig. \ref{fig:sim1}, we plot the radar energy efficiency (REE) versus the communication energy efficiency (CEE).
Specifically, the CEE and REE are defined as $\text{CEE} = \sum_{u\in\mathcal{U}}{\text{Rate}_u}/P_{\rm TOL}$ and $\text{REE} = {\text{RMI}}/P_{\rm TOL}$, respectively.
It is observed in the right side of Fig. \ref{fig:sim1} that FC has the worst energy performance, while SPS-based architectures have the best energy performance.
Besides, the dynamic architectures are slightly better than conventional fixed architectures. 
From Fig. \ref{fig:sim1}, we conclude that the proposed Dym-DPS can provide the best trade-off between performance and energy efficiency.

In Fig. \ref{fig:sim4}, we investigate the impact of the number of transmit antennas $N_{\rm T}$ with QoS threshold $\Gamma = 5$.
From left side of Fig. \ref{fig:sim4}, we observe that the RMI will improve with the number of antennas.
This is because increasing the number of antennas can offer more antenna diversity and thus larger beamforming gain.
Notably, the RMI performance achieved by different architectures tends to saturate with increasing number of antennas.
Furthermore, the proposed Dym-DPS architecture always outperforms other benchmark architectures.
From right side of Fig. \ref{fig:sim4}, we observe that the REE decreases with increasing number of antennas due to the increasing energy consumption induced by more antennas, PSs, and switches.
Among all architectures considered, the REE achieved by FC performs the worst, while the Dym-SPS has the best REE.
In addition, with the growth of the number of antennas, the gap between dynamic and conventional fixed architectures becomes smaller.
This is because further increasing the number of antennas does not result in significant performance gains but instead leads to a substantial increase in system power consumption.
Therefore, it is important to appropriately choose the number of antennas in practice to balance performance and energy efficiency.

\begin{figure}
    \centering
	\includegraphics[width=0.65\linewidth]{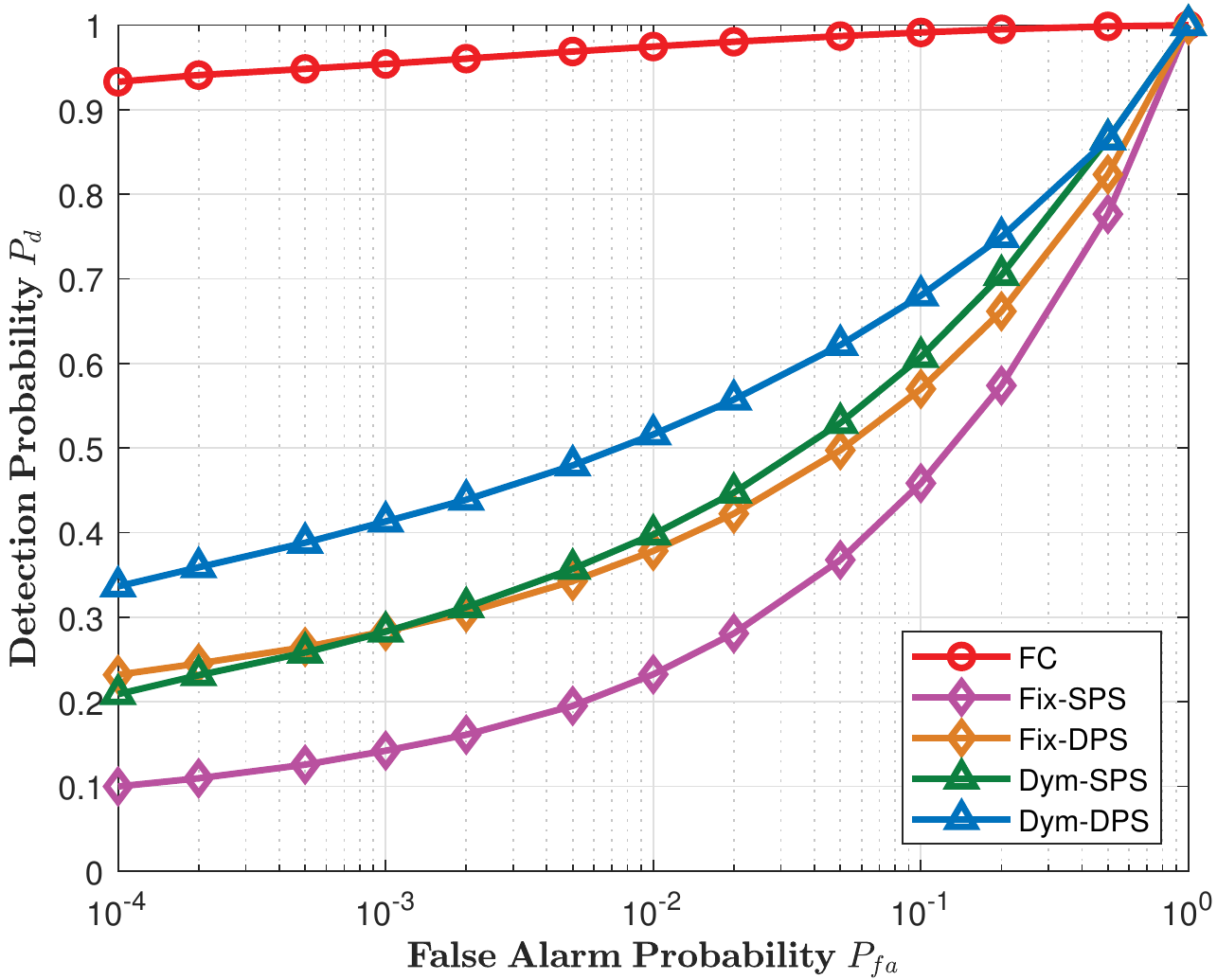}
    \vspace{-1em}
	\caption{Comparisons of ROC with QoS threshold $\Gamma = 5$.}
	\label{fig:sim2}
    \vspace{-1.em}
\end{figure}

In Fig. \ref{fig:sim2}, we depict the radar receiver operating characteristic (ROC) curves for different architectures with $\Gamma = 5$.
As can be observed from Fig. \ref{fig:sim2}, the probability of detection $P_d$ of all architectures increases with the probability of false alarm $P_{fa}$.
As expected, the proposed Dym-DPS architecture achieves the best detection performance among all sub-connected ABFs.
Combining Figs. \ref{fig:sim1} with \ref{fig:sim2}, we can conclude that higher RMI yields better detection performance.

\begin{figure}
    \centering
	\includegraphics[width=0.65\linewidth]{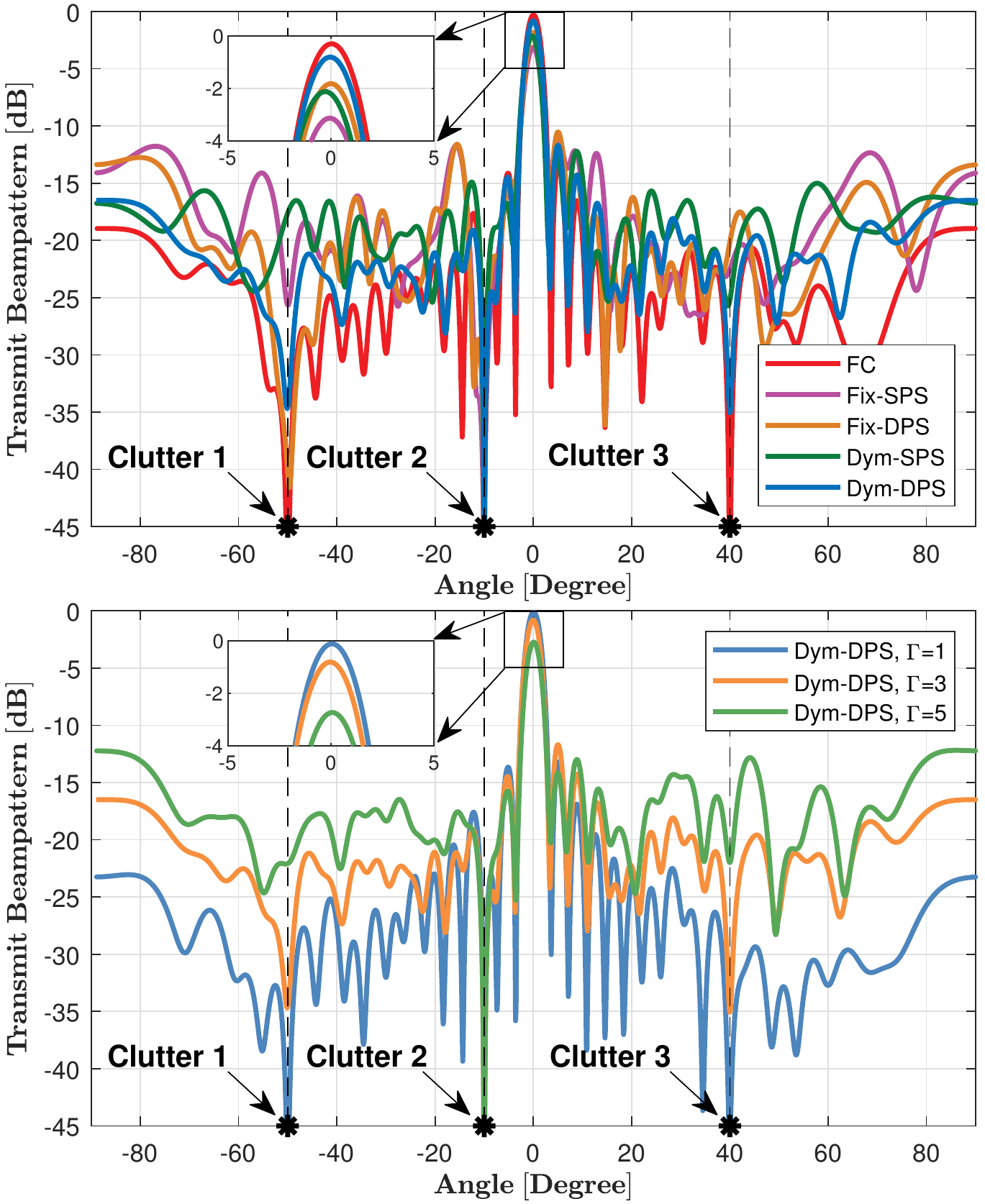}
    \vspace{-1em}
	\caption{Comparisons of transmit beampattern. Top: different architecture with $\Gamma$=3. Bottom: Dym-DPS architecture with different QoS threshold $\Gamma$.}
	\label{fig:sim6}
   \vspace{0.5 em}
\end{figure}

In Fig. \ref{fig:sim6}, we show the transmit beampattern behaviors of the proposed DFRC.
From the top of Fig. \ref{fig:sim6}, we observe that among all sub-connected architectures, the proposed Dym-DPS can achieve sharp nulls at clutter directions and concentrate more energy towards the target direction, which guarantees a high MI at target directions while suppressing clutters.
Moreover, from the bottom of Fig. \ref{fig:sim6}, we find that the smaller the communication threshold $\Gamma$, the better the peak-to-side lobe ratio of the beampattern will be achieved, further confirming the trade-off between radar and communication.

\section{Conclusions}

In this paper, the hybrid beamforming design of the DFRC BS system with dynamically sub-connected HBF architectures is studied.
Specifically, we consider the hybrid beamforming design to maximize the RMI subject to communication QoS, power budget, and ABF hardware constraints.
An effective algorithm for solving the resulting complicated non-convex optimization problem by the AO and MM methods is presented.
Simulation results show that the dynamically sub-connected architecture can accelerate the convergence speed and outperform conventional sub-connected HBF architectures.
In the future, it is interesting to extend this novel dynamic HBF architecture to the robust design for DFRC, wideband OFDM DFRC, as well as distributed DFRC scenarios.

\appendices

\footnotesize
\bibliographystyle{IEEEtran}
\bibliography{IEEEabrv,L_ref.bib}

\end{document}